# Application of Long-short Term Memory (LSTM) Model for Forecasting NOx Emission in Pohang Area

Sangdeok Lee (kamgissang@unist.ac.kr), MinChung Kim (mckim@unist.ac.kr)

국문초록

일산화질소와 이산화질소로 명명된 질소산화물(NOx)의 배출은 환경 및 건강의 주요 관심사이다. 이러한 기후 위기에 대응하기 위해 대한민국 정부는 질소산화물의 배출 규제를 강화해 왔다. 정확한 질소산화물 예측 모델은 기업이 질소산화물 배출 할당량을 준수하고 비용 절감을 달성하는 데 도움이 될 수 있다. 본 연구는 대기오염 문제가 심각한 한국의 중공업 도시인 포항의 질소산화물 배출량을 예측하는 모델을 개발하는 데 초점을 맞췄다. 확률 회귀를 이용한 결측치 추정과 함께 LSTM(Long-short term memory) 모델링을 적용하여 NOx 배출량을 예측하였다. LSTM 모델을 실행하는 데 있어 두 가지 파라미터(time windows 와 learning rates)를 조절하였으며, LSTM 에서 널리 사용되는 최적화 방법의 하나인 Adam optimizer 를 선택하였다. 중요한 평가 기준인 평균 절대 오차(MASE)가 1 미만으로써, 적용된 모델이 허용 가능한 예측 성능을 달성하였다. 이는 해당 모델을 적용하면 naive prediction(마지막으로 관측된 데이터 포인트를 기반으로 예측하는 모델)에 비해 미래 질소산화물 배출량을 예측하는 데 있어 더 나은 성능을 발휘한다는 것을 의미한다.

**Abstract**

Emissions of nitric oxide and nitrogen dioxide, which are named as NOx, are a major environmental and health concern. To react to the climate crisis, the South Korean government has strengthened NOx emission regulations. An accurate NOx prediction model can help companies to meet their NOx emission quotas and achieve cost savings. This study focuses on developing a model which forecasts the amount of NOx emissions in Pohang, a heavy industrial city in South Korea with serious air pollution problems. In this study, the Long-short term memory (LSTM) modeling is applied to predict the amount of NOx emissions, with missing data imputation using stochastic regression. Two parameters (i.e., time windows and learning rates) necessary to run the LSTM model are tested and selected using the Adam optimizer—one of the popular optimization methods in LSTM. I found that the model that I applied achieved the acceptable prediction performance since its Mean Absolute Scaled Error (MASE)—the most important evaluation criterion—is less than 1. This means that applying the model that I developed in predicting future NOx emissions will perform better than a naive prediction, a model that simply predicts them based on the last observed data point.

# Contents

1. **Introduction**

Climate change is a global environmental issue. Carbon neutrality has been a global priority since the Kyoto Protocol in 1997, aiming for net zero by 2050 and limiting the increase of global temperature beyond 1.5°C under the Paris Agreement of 2015. [1] The US's withdrawal from the agreement under President Trump caused a setback, [2] but President Biden restored US participation. [3] Major companies such as Google, Apple, Amazon, and GM are also committing to carbon neutrality, with Google claiming to have achieved it since 2007 and others aiming for it by 2030 or 2040. [4]

NOx (i.e., nitric oxide and nitrogen dioxide) emissions are a major concern due to their environmental impact and health risks. They contribute significantly to regional and local air pollution and can cause acidification of soil and water, leading to decreased crop yields and forest growth. Additionally, NOx can produce ground-level ozone, which is toxic. [5] The World Health Organization (WHO) recommends forecasting air pollution levels to reduce the associated health risks. [6]

South Korea responded to combat this climate emergency. South Korea ratified the Paris Agreement on November 3, 2016. As a signatory to the agreement [7], this has led to the implementation of the "2030 National Greenhouse Gas Reduction Goals (NDC-Nationally Determined Contribution)". This policy is expected to greatly reduce the emissions of air pollutants like SOx and NOx. [8] NOx is primarily emitted from fossil fuel combustion and other sources, such as biomass burning and microbial N cycles in soils and animal waste. [9] One way to reduce NOx emissions is to transit from fossil fuels to renewable energy sources. South Korea's "Green New Deal" is a prime example of this approach, as it aims to achieve carbon neutrality by 2050 by shifting away from fossil fuels, promoting the development of a hydrogen economy, and expanding renewable energy sources. [10] The regulation of diminishing NOx emissions is getting stronger over the years.

For example, until 2021, the Ministry of Environment will charge businesses KRW 1,810 per kilogram for emissions above 50% of the permitted limit, and KRW 2,130 per kilogram for emissions above 30% of the permitted limit from 2022. [11] A number of South Korea's government departments makes efforts to reduce the emissions of toxic gases such as NOx. In 2021, the Ministry of Environment, the Ministry of SMEs and Startups, the Ministry of Trade, Industry and Energy, the Korea Advanced Institute of Production Technology, and the Korea Environment Agency implemented a Clean Factory Construction Project for 300 workplaces with a budget of 20.3 billion won. [12] This was an increase from the 31 billion won in 2020 that supported 98 workplaces. The policy is set to continue in 2022. [12] The Statista reported that South Korea emitted about 1.09 million tons of nitrogen oxides in 2019, which decreased from around 1.25 million tons in 2016. According to another study, the NOx emission standard for Korean cement manufacturers has decreased steadily. It was 270 ppm or less for kilns installed prior to January 31, 2007, 200 ppm or less for kilns installed between February 1, 2007, and December 31, 2014, and 80 ppm or less for installations implemented on or after January 1, 2015. [13]

However, despite this commitment, South Korea in 2020 still relied on coal for approximately 40% of its electricity generation, with renewables accounting for less than 6%. [14] NOx has been found in several cities in Korea. According to a study conducted in 2014, the cities of Gyeongju, Pohang, and Ulsan were found to have particularly high levels of NOx. These elevated levels could potentially lead to health and environmental problems in those areas. [15]

Among those cities, Pohang is an important industrial city in South Korea. According to the steel industry and POSCO, Pohang Steel produced 16.85 million tons of crude steel last year, accounting for 35% of the country's total crude steel production. [16] POSCO Chemical is planning to build its third factory for cathode materials in South Korea by 2025 in this city, which is estimated to cost 600 billion won (about 445 million euros). [17] It is essential to keep the factory in a pristine condition to guarantee the production of high-quality goods. Unclean and dusty environments can result in lower productivity, diminished product quality, and increased manufacturing expenses. By keeping the factory clean, productivity can be improved, and the environmental impact can be reduced. [18] Nevertheless, it is estimated that POSCO's factory will emit the most air pollutants in 2022, ranking 1-2 in the nation. The POSCO's factory in Pohang emitted 15,436 tons of NOx. [19] POSCO has received several administrative penalties for exceeding air pollutant emission standards. [20] Thus, this research will focus on developing a method of forecasting NOx levels using data from a firm located in the Pohang area for the purpose of having further control over the environmental indicators.

## 2. Literature review

### a. Literature review of the models forecasting NOx

Forecasting is a critical element in an environmental study. Several studies used the data-driven method as an effective way to predict NOx emissions. Various algorithms, such as statistical regressions [21][22], support vector machines [23][24][25][26], deep learning[1] [27][28][29], and artificial neural networks (ANNs) [30][31][32][33][34][35][36][37][38][39], have been used to predict NOx emissions.

Studies comparing forecasting models specifically about NOx were also done in previous studies. César Bouças et al. (2020) compared several forecasting models (e.g., Moving Average, Linear Regression, and Long Short-Term Memory [LSTM]) in terms of forecast error. They showed that using LSTM gives the best forecast performance regarding root mean squared error (RMSE) and mean absolute percentage error (MAPE), and then the performances of linear regression and moving average are followed. [40] Thus, in this project, I am going to use LSTM based on performance test results done by César Bouças et al. (2020).

---

[1] Deep learning and ANN are both machine learning methods. The key difference between ANN and deep learning is the number of hidden layers. ANN typically have one or two hidden layers, while deep learning networks can have dozens or even hundreds of hidden layers.

### b. History of the development of the LSTM method

In order to explain the essence of the LSTM method in forecasting the amount of NOx emissions, it is necessary to understand the ANN and Recurrent Neural Network (RNN) methods. ANN was first introduced in 1943 by Warren McCulloch and Walter Pitts. [41] The paper was published in the Bulletin of Mathematical Biophysics. It was a paper to deliver a mathematical approach to theoretical neurophysiology to explain the nervous system as a net of neurons, each having a soma and an axon. In short, it was trying to create a mathematical model mimicking a biological nervous system.

However, there is a limitation to ANN itself. ANN cannot capture long-term dependencies as the number of data points is fixed. This trait is acceptable for most classification problems, but not for time series data. RNN was later developed in 1985 by Rumelhart et al. [42] RNN is considered a type of ANN. One of the benefits of using RNN over other ANNs is its ability to represent sequences of data, allowing for the assumption that each individual sample depends on preceding ones. In other words, RNNs can utilize previous information and use it to make decisions or predictions. [43]

Hochreiter and Schmidhuber introduced LSTM in 1997 as an improved version of RNN. [44] The limitation of RNN itself was the 'vanishing gradient problem' where information gets lost when long-term dependencies get lost exponentially. LSTM overcomes this problem by adding the concept of 'gates'. [43] The gates multiply the importance of data from 0 to 1. The important data will survive over time by multiplying the value of the data by nearly 1, and the less important data will vanish by multiplying the value of the data by nearly 0.

Based on the literature reviews above, it is widely considered that ANN is a reliable method for performing forecasting for NOx. Additionally, among ANNs, LSTM is considered to perform well under time series data.

Therefore, in this study, LSTM will be used to forecast Pohang's NOx level, as it is widely used as a reliable method in other studies.

## 3. Methodology

### a. Data description

I obtained the dataset used for this study from the Korea Environment Corporation via the Korean Open Data Portal. [45] The dataset includes the amount of NOx emissions every 30 minutes on each day from October 2021 to April 2023, which has been recorded by a chimney-automated measuring device located in the Pohang area. [45] In addition to the NOx emission variations, the dataset includes different variables, as shown in Table 1.

Table 1. Variables in the dataset

| Variable | Meaning |
|----------|---------|
| area_nm | Factory location |
| measure_dt | date/time of the measurement |
| TSP | Amount of dust emission |
| SOx | Amount of Sulpher Oxide emission |
| NOx | Amount of Nitric Oxide emission |

### b. Data imputation

As 14,046 data points are missing, the data needs to be imputed. This step is important as missing data can negatively impact model's performance. If imputation is not done correctly, it could result in inaccurate predictions. [46] There are omitted observations (14,046 missing observations) probably due to the maintenance or upgrade of the chimney-automated measuring device. Thus, the number of NOx emission observations in the dataset is 41,924.

The type of imputation that is done is stochastic regression imputation. The method utilizes ordinary least squares (OLS) regression to anticipate missing values and then adds a residual term that is normally distributed to each predicted value. By adding a random error term to the predicted value, this approach can more accurately replicate the correlation between variables. [47]

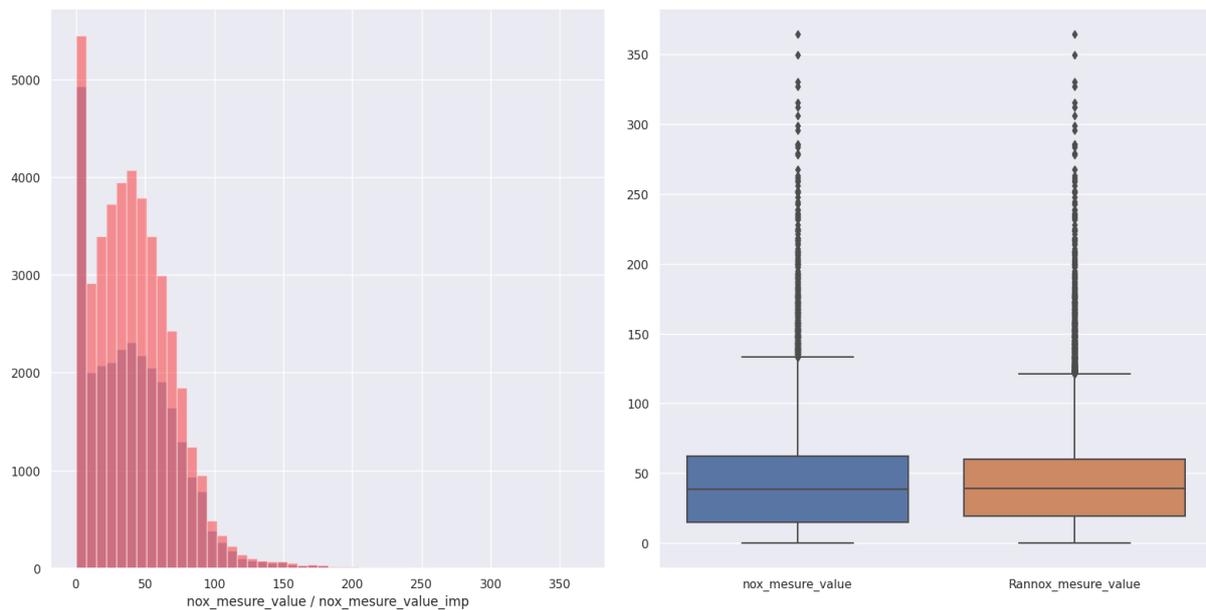

[Fig1] Comparing NOx values before (blue) and after (orange) the imputation.

There are two overlapping histograms on the left in [Fig1] where the blue histogram shows the distribution of the original data, and the transparent orange color histogram shows the imputed data. On the right of [Fig1], the blue boxplot on the left is the original data, and the orange boxplot is the data after the stochastic regression imputation. It indicates that the result of the stochastic regression imputation shows a similar statistical variance and distribution. Thus, it is likely to show that an inaccurate prediction has diminished.

### c. Estimation with LSTM

I developed the programming code using Python on a Linux operating system based on deep learning libraries such as Keras [48] and sci-kit-learn [49]. The model utilizes the Adam optimizer[2] for LSTM. [50] Adam is chosen due to its computational efficiency and superior performance in comparison with different stochastic optimization methods. [51][52] The method is trained for 100 epochs [53], and the batch size is set to 64, a common value that is a power of 2. [54] The study conducted tests using various configurations of LSTM hidden neurons, including 16, 32, and 64. The results on the accuracies showed that the LSTM model with 64 hidden neurons had the best performance. Hence, the study employed a one-layer LSTM architecture with 64 hidden neurons to conduct a 30-minute prediction.

### d. Data setting regarding forecasting

To forecast NOx emissions at time t, I consider the historical NOx emission data from time window[3] 9 (i.e., t-9; 30mins*9; 4 hours 30 minutes before the forecast time) to time window 2 (i.e., t-2; 30mins*2; 1 hour before the forecast time) as the possible range of the historical data for forecasting. I check the accuracy and figure out which time window I will use for forecasting. For example, if time window 7 gives the highest accuracy, I will choose time window 7 for forecasting.

---

[2] A term for "Adaptive Moment Estimation". This optimization is a widely used in Keras' LSTM model because, according to Kingma et al. (2014), It possesses computational efficiency, demands minimal memory, remains unaffected by diagonal rescaling of gradients, and is particularly suitable for addressing data or parameter-intensive problems. [55][56]
[3] Here, time window refers to the point of time in the historical data used for the forecasting.

A hyperparameter "learning rate" [4] is also adjusted to find the best model based on accuracy. The parameters that are evaluated are $-10^1$, $-10^2$, $-10^3$, $-10^4$, $-10^5$, $-10^6$. This is because a high learning rate can result in the optimization process overlooking the optimal point, leading to a decrease in the accuracy of the model's predictions. On the other hand, a lower learning rate may cause the model to take a longer time to converge. [31]

Following Kang et al. [30] and Tan et al.'s [31] study, I considered the first 29,347 (i.e., the first 70% of the total observations in time order) as the training dataset while the second 12,577 (i.e., the last 30% of the total observations in time order) as the test dataset.

### e. Forecasting evaluation metrics

To test the performance of the forecasting model that I propose, I use the three metrics: Mean Absolute Scaled Error (MASE) [5], Root Means Square Error (RMSE), and Coefficient of Determination ($R^2$) (hereinafter as 'accuracy') [6]. Formulas for MASE, RMSE and accuracy are

$$MASE = \frac{1}{n}\sum_{t=1}^{n}\frac{|y_t - y_t'|}{\frac{1}{n-1}\sum_{i=2}^{n}|y_t - y_t' - 1|}$$

$$RMSE = \sqrt{\frac{1}{n}\sum_{t=1}^{n}(y_t - y_t')^2}$$

$$Coefficient\ of\ Determination\ (R^2)[Accuracy] = 1 - \frac{MSE}{\sigma^2(y_t)} = 1 - \frac{RMSE^2}{\sigma^2(y_t)}$$

---

[4] During training, the model adjusts its weights or parameters to become more accurate. The learning rate determines the size of each adjustment made during this process.

[5] Mean Absolute Percentage Error (MAPE) was considered as it is a widely used parameter. However, MAPE had a critical limitation that can lead to misinterpretation if observations contain zero or near zero. [57] The dataset contains several observations that are zero or near zero. Because of the formula, the division of zero will lead to the outcome to have a tendency of infinity. Thus, MASE is used as an alternative.

[6] This accuracy measure represents how well the model's predictions match the actual values, relative to the variance of the actual values. A higher accuracy value indicates that the model's predictions are closer to the actual values. It is a measure of how accurately a statistical model can predict an outcome. [58]

where $y_t$ is the measured value at time $t$; $y_t{'}$ the predicted value at time $t$, $n$ as the number of observations, and $\sigma^2(y_t)$ as variance of $y_t$.

## 4. Empirical results

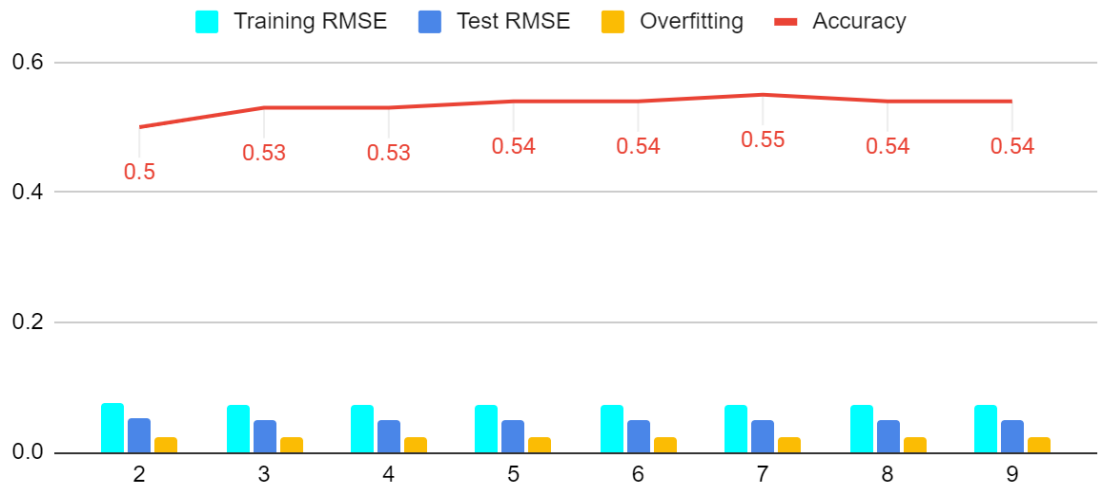

[Fig2] The result of RMSE and accuracy over the time window from 2 to 9 when learning rate = 0.0001 ($-10^4$)

To determine the time window for the historical data, I checked the accuracy for time window 2, 3,…, 9. According to [Fig2], time window = 7 gives the highest accuracy.

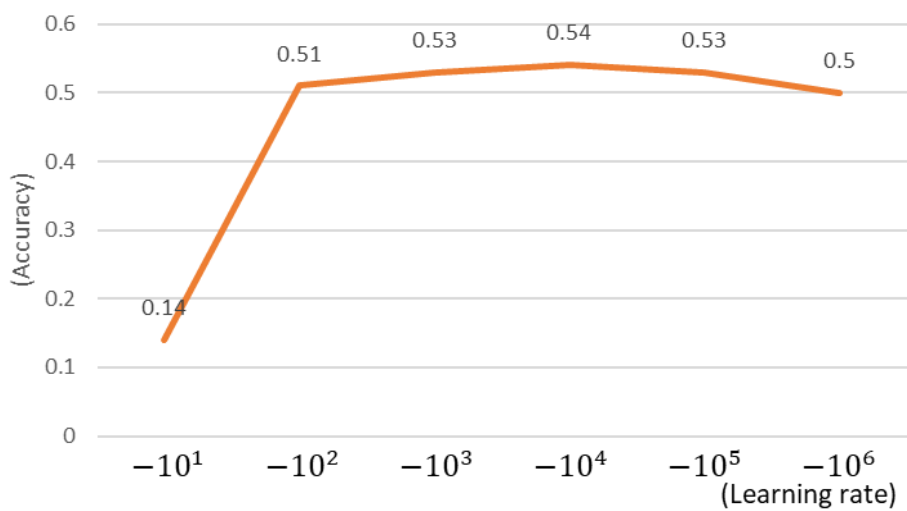

[Fig3] Accuracy based on Learning rate in a time window = 7

Learning rate is another key information that I need to choose, which determines the model performance. To choose the learning rate, six parameters are evaluated: -10$^1$, -10$^2$, - $10^3$, $-10^4$, $-10^5$, $-10^6$. Among the six values, the learning rate of 0.0001 ($-10^4$) is selected since it provides the best prediction accuracy as shown in [Fig3]. Thus, time window = 7 and a learning rate of 0.0001 ($-10^4$) are selected to run the LSTM model.

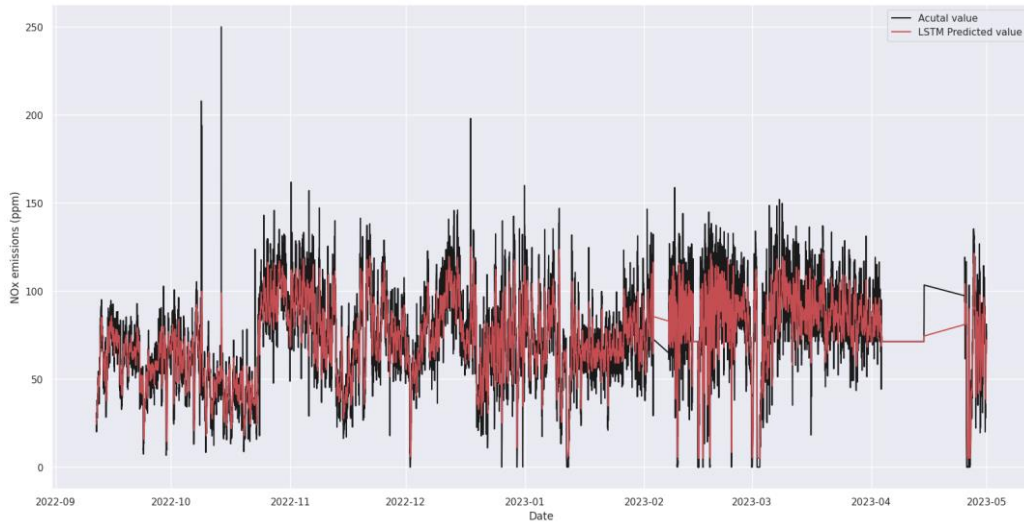

[Fig4] A comparison of the actual data (black) and the predicted data (red) from the test dataset in a time window = 7

To check the performance of the model I developed, I used the 'accuracy' measure in the model outputs. The accuracy means the percentage of observations in the test dataset which were precisely predicted by using the proposed model. Before interpreting the model performance metric—accuracy, it is necessary to check whether the model is overfitted[7]. One way of checking overfitting is the difference between the RMSE obtained by estimating the training dataset and the RMSE obtained by estimating the test dataset. I find that the RMSE of the training dataset is 0.0731 and the RMSE of the test dataset is 0.0499, whose difference is +0.0232, suggesting that the model is not overfitting as the training RMSE is higher than the testing RMSE which are implemented and interpreted identically by Derraz et al. (2023) and Liu et al. (2022)[8] [59][60]

---

[7] It means a machine learning model becomes too specialized in learning from a training dataset to the extent that it struggles to perform well on other new or unseen datasets like the test dataset. For example, when a model trains for too long on training data, it can start to learn the irrelevant information, or "noise", within the dataset. This will be a problem as a overfitted model cannot be generally used in other dataset

[8] The model is not overfitted if the training RMSE is higher than the testing RMSE.

I find that the accuracy is 55% with time window = 7, learning rate = 0.0001, and epoch = 100. The paper primarily focuses on the prediction results for the testing dataset, where the model achieves an accuracy of 55%. These results indicate that the model is accurate. Especially, the MASE value of 0.88 on the test set indicates that the model is better than a naive model, which simply predicts NOx emissions based on the observations in the last time window. [Fig4] shows the model performance visually where the forecast model output (red) follows the tendency of the actual data (black). Thus, the results prove that it is worthwhile to apply LSTM model that I developed in predicting NOx emissions at least in the Pohang area.

## 5. Limitations and further studies

The LSTM model that I developed for NOx emission prediction performs well. Nonetheless, I hope to achieve better accuracy. For example, I have made a forecasting model of SOx emission based on the SOx observations from the same dataset. It is modeled under identical parameters (identical number of observations, identical imputation method, and identical forecast modeling using LSTM at time window = 7, learning rate of 0.0001, and epoch of 100). The accuracy of the forecasting model of SOx emission is 71%, which is an improvement of the NOx emission model's 55%. It is more interesting since the MASE value of this model is 0.96, higher than the forecasting model of NOx emission.

It is extremely difficult to find out the reasoning behind the model's performance based on LSTM as LSTM is a black box model. [61][62] Nevertheless, an educated speculation can be made.

For instance, Takeuchi et al. (2018) suggested that the higher the data entropy[9], the lower the accuracy of the ANN model. [63] The NOx total dataset's data entropy is 8.587, and a random selection of 2,000 observations from the NOx dataset is 7.09.[10] A bitcoin dataset of the last 2,000 days is 7.60. [64] Thus, it may indicate that making forecast modeling based on the dataset of NOx is as difficult as making a forecast model for bitcoin price predictions as the entropy level of the NOx dataset is close to the bitcoin dataset.

If more data are available, more complex versions of LSTM models ─ that require additional computational time [65] ─ can be used in order to enhance model performance. For example, Wang et al. (2020) [66] conducted a study of NOx prediction based on LSTM using 10,000 data points. Two years later, Wang et al. (2022) conducted another study of NOx prediction with 100,000 data points from the same data source and used the CEEMDAN-AM-LSTM model. Based on Wang et al.'s (2022) findings, the CEEMDAN-AM-LSTM model shows an improved prediction result over the existing LSTM model based on RMSE. [67] When more extensive data, future research can apply the model developed by Wang et al. (2022) in the prediction of the emission of NOx probably with more computational time.

Additionally, based on Yu et al. (2020)'s findings, the model's accuracy may be enhanced if the model's epochs or iterations are increased under a lower learning rate. [68] Thus, future studies may improve the forecast performance by increasing epochs under lower learning rates.

## 6. Discussion

---

[9] Data entropy is a measure of the uncertainty or randomness of a dataset. It is a statistical concept that describes the amount of information contained in the data.

[10] The reason to randomly select 2,000 observations in NOx dataset is to match the number of observations in the bitcoin dataset; and ensure a fair comparison.

In this study, I applied the LSTM model in forecasting NOx emissions with the data of NOx emission amounts every 30 minutes from October 2021 to April 2023 in the Pohang area. To do that, first I imputed missing observations since the dataset includes many missing observations possibly due to device updates or malfunctioning. I then estimated the key parameters in the LSTM model which are time window and learning rate. Using the estimated parameters, I predicted NOx emission amounts in a test dataset and found an acceptable level of accuracy which indicates the value of applying the LSTM model in predicting NOx emissions in the Pohang area.

The use of the forecast of NOx can benefit for factories and researchers of interest.

First, by accurately predicting NOx emissions, factories that have power plants, industrial boilers, cement kilns, and turbines[11] can reduce their environmental-related operating costs and comply with the Emissions Trading System (ETS) for carbon emissions. There are many cases where forecasting models prevent cost expenditures. For example, Zhang et al. (2022) indicated that precise load forecasting[12] can be a useful strategy to minimize the expenses associated with power system operations. [69] The Boston Consulting Group has highlighted the potential benefits of implementing an AI-based method to measure and mitigate emissions, which can lead to increased revenue and cost savings. [71]

Second, these forecasts can also be used in environmental models for air pollution, taking into account factors such as weather and wind. For example, Chang et al. (2022) presented a hybrid model that utilizes weather conditions to predict and minimize multiple air pollutants. [72] Wang et al. (2022) developed an aggregated model based on Long Short-Term Memory (LSTM) for air pollution forecasting. [73]

---

[11] It is known that factories with power plants, industrial boilers, cement kilns, and turbines are highly likely to emit NOx. [70]
[12] Forecasting energy demand and supplies.

Lastly, the Pohang area's public health can be protected if future studies identify the sources that causes NOx emission and implement their study in actual factory operations. For example, according to research conducted by the US Environmental Protection Agency (EPA), the decrease in fine particle pollution in US cities from 1980 to 2000 resulted in an increase in average life expectancy at birth of roughly seven months. This reduction in air pollution was due to measures such as the implementation of the Clean Air Act, which regulates emissions from factories using modern pollution control technology based on accurate predictions of such emissions. Therefore, through further development of the prediction model of toxic gases like NOx in the factory level, and its application to Pohang area, public health conditions in the Pohang area can be enhanced. [15][74]

In conclusion, the prediction model can be beneficial in threefold: reduction of environmental-related operating costs, further usage of environmental models for air pollution, and contribution to addressing local health issues by developing the prediction model and implementing it in factories' emission control operations.

**Appendix 1** - **The structure of LSTM**

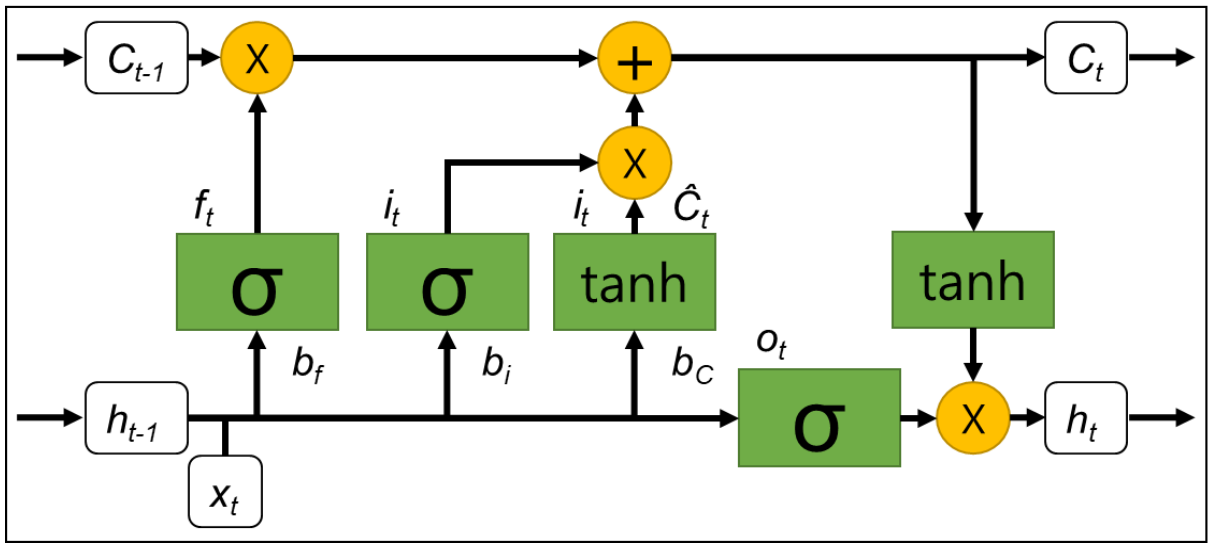

In detail, the computational process of LSTM as follows:

Step 1, determine which information should be retained or discarded in the forget gate. The following equation is

$f_t = \sigma(W_f \cdot [h_{t-1}, x_t] + b_f)$

where

$f_t$ is the forget gate at time step t

$\sigma$ is the sigmoid function ($\sigma = \frac{1}{1+e^{-x}}$)

$W_f$ is the weight matrix for the forget gate

$h_{t-1}$ is the previous hidden state

$x_t$ is the input at time step t

$b_f$ is the bias term for the forget gate

The outcome of the value is between 0 and 1. It will preserve the information if the value is closer to 1, and discard if the value is closer to 0.

The input layer of LSTM is responsible for determining what new information needs to be added to the cell state.

Step 2, the previous hidden state and the current input are fed into a sigmoid function:

$i_t = \sigma(W_i. \ [h_{t-1}, x_t] + b_i)$

where

$i_t$ is the input gate vector at time step t

$\sigma$ is the sigmoid function ($\sigma = \frac{1}{1+e^{-x}}$)

$W_i$ is the weight matrices of the input gate

$h_{t-1}$ is the previous hidden state

$x_t$ is the current input at time step t

$b_i$ is the bias matrices of the input gate

Step 3, the hidden state and the current input are again passed through the tanh function in order to regulate the amount of information that is added to the cell state.

$C_t = tanh(W_C. \ [h_{t-1}, x_t] + b_C)$

where

$C_t$ is the candidate cell state at time t

$W_C$ is the weight matrices of the cell state

$h_{t-1}$ is the previous hidden state

$x_t$ is the current input t time step t

$b_C$ is the bias matrices of the cell state

Step 4, update the cell state of the memory cells.

$C_t = f_t \ *C_{t-1} + i_t \ *C_t$

Step 5, the output is obtained along with the new hidden state.

$o_t = \sigma(W_o. \ [h_{t-1}, x_t] + b_o)$

$h_t = o_t *tanh(C_t)$

[45]